\documentstyle[11pt,fleqn]{article}   
\def\maketitle{\thispagestyle{empty}\setcounter{page}0\newpage
                \renewcommand{\thefootnote}{\arabic{footnote}}
                  \setcounter{footnote}0}
\renewcommand{\thanks}[1]{\renewcommand{\thefootnote}{\fnsymbol{footnote}}
               \footnote{#1}\renewcommand{\thefootnote}{\arabic{footnote}}}
\newcommand{\preprint}[1]{\hfill{\sl preprint - #1}\par\bigskip\par\rm}
\renewcommand{\title}[1]{\begin{center}\Large\bf #1\end{center}\rm\par\bigskip}
\renewcommand{\author}[1]{\begin{center}\Large #1\end{center}}
\newcommand{\address}[1]{\begin{center}\large #1\end{center}}

\def\dinfn{\smallskip Dipartimento di Fisica, Universit\`a di Trento\\ 
                           and Istituto Nazionale di Fisica Nucleare,\\
                                   Gruppo Collegato di Trento, Italia}

\def\Idinfn{\address{\dinfn}}
\newcommand{\email}[1]{e-mail: \sl #1@science.unitn.it\rm}
\newcommand{\femail}[1]{\thanks{\email{#1}}}
\newcommand{\pacs}[1]{\smallskip\noindent{\sl PACS numbers:
                       \hspace{0.3cm}#1}\par\bigskip\rm}
\def\babs{\hrule\par\begin{description}\item{Abstract: }\it} 
\def\eabs{\par\end{description}\hrule\par\medskip\rm}
\renewcommand{\date}[1]{\par\bigskip\par\sl\hfill #1\par\medskip\par\rm}
 
\renewcommand{\vec}[1]{{\bf #1}}       
\def\M{{\cal M}}                       
\newcommand{\ca}[1]{{\cal #1}}         
\def\hs{\qquad}               
\def\nn{\nonumber}            
\def\beq{\begin{eqnarray}}    
\def\eeq{\end{eqnarray}}      
\def\at{\left(}               
\def\aq{\left[}               
\def\ag{\left\{}              
\def\cp{\right.}              
\def\ct{\right)}              
\def\cq{\right]}              
\def\R{{\hbox{{\rm I}\kern-.2em\hbox{\rm R}}}}   
\def\H{{\hbox{{\rm I}\kern-.2em\hbox{\rm H}}}}   
\def\N{{\hbox{{\rm I}\kern-.2em\hbox{\rm N}}}}   
\def\C{{\ \hbox{{\rm I}\kern-.6em\hbox{\bf C}}}} 
\def\Z{{\hbox{{\rm Z}\kern-.4em\hbox{\rm Z}}}}   
\def\ii{\infty}                                  
\def\X{\times\,}                                 
\newcommand{\fr}[2]{\mbox{$\frac{#1}{#2}$}}      
\def\Tr{\mathop{\rm Tr}\nolimits}                  
\renewcommand{\Re}{\mathop{\rm Re}\nolimits}       
\def\lap{\Delta}                                   
\def\al{\alpha}
\def\be{\beta}

\def\ep{\varepsilon}
\def\ze{\zeta}

\def\la{\lambda}

\def\Ga{\Gamma}

\def\La{\Lambda}

\begin{document}

\preprint{UTF-389}
\title{Zeta-function on the generalised cone}

\author{Guido Cognola\femail{cognola} and Sergio Zerbini\femail{zerbini}}
\Idinfn

\date{september 1996}

\babs
The analytic properties of the $\zeta$-function for a 
Laplace operator on a generalised cone $\R^+\X\M^N$ are studied 
in some detail using the Cheeger's approach and explicit expressions are given.
In the compact case, the $\zeta$-function of the Laplace operator  
turns out to be singular at the origin. As a result, strictly speaking, 
the  $\zeta$-function regularisation does not ``regularise'' and a further 
subtraction is required for the related one-loop effective potential.  
\eabs


Manifolds  with conical singularities attracted the 
interest of some physicist since the beginning of the century with the 
works of Sommerfeld, 
but it was in the last decades that they become  popular among
all physicists working on space-times with horizons.
The reason is due to the fact that in such kind of space-times there 
is a natural equilibrium temperature, the Hawking temperature, which, 
within the Euclidean approach, can be easily computed by imposing the 
absence of the conical singularity \cite{gibb77-15-2752}, 
i.e. by requiring the space-time  to be a smooth manifold. 
A lot of work has been done in this direction, mainly concerning the 
pure cone, where heat kernel 
\cite{call94-333-55,suss94-50-2700,kaba95-453-281,empa95-51-5716,solo95-51-618,furs95-10-649} and 
$\zeta$-function\cite{cogn95-12-1927,byts96-458-267,zerb96-54-2699} 
and their applications to physics have been studied in some 
detail. The generalised cone has been investigated for the first 
time in a seminal paper by Cheeger \cite{chee83-18-575}, where the interested 
reader can find the general properties concerning the heat kernel and 
$\zeta$-function related to the Laplace opearator on functions
and more recently by Bordag, Dowker and 
Kirsten\cite{bord96u-82,dowk96u-189}, 
where the generalisation to generic $p-$forms has also been carried out.
Here we derive explicit analytic expressions for the  
$\zeta$-function of a Laplace operator acting on functions 
(massless scalar fields) in a generalised cone with an arbitrary 
smooth base, following the Cheeger's approach 
\cite{chee83-18-575}.

To start with we remind that the $D=N+1$
dimensional generalised cone $\M^D=\R^+\X\M^N$ has local properties 
described by the metric
\beq
ds^2=dr^2+r^2 d\sigma^2_N
\:,\label{cone}\eeq
where $d\sigma^2_N$ is the metric of the compact smooth manifold 
$\M^N$, with or without boundary (the base). 
Let us denote $x=(r,\vec \tau)\in\M^D$, $\vec \tau\in M^N$. 
For example, for the pure cone with deficit angle $2\pi-\be$,
 $0<\tau<\be$, $\be$ being a parameter which takes 
the conical singularity into account. If $\be=2\pi$ then $\M^1\equiv 
S^1$ and in this case $\M^2$ is a smooth manifold.
If $r\in R^+$, the generalised cone is non-compact 
and the spectral properties of the Laplace operator 
on $\M^D=\R^+\X\M^N$  are well known, the spectrum is continuous 
and a complete set of normalised eigenfunctions
of the operator (negative Laplacian)
\beq 
L_D=-\frac{\partial^2}{\partial r^2}
-\frac{N}{r^2}\frac{\partial}{\partial r}
-\frac{1}{r^2}\lap_N\:,
\eeq
is easily found to be  
\begin{equation}
\psi_{\la\al}(r,\vec \tau)= r^{\frac{1-N}2}
J_{\nu_\al}(\la r)\:\phi_\al(\vec \tau)
\:.\end{equation}
Here $\la^2$ ($\la\geq0$) is the continuous eigenvalue corresponding to 
$\psi_{\la\al}$, while $J_{\nu_\al}$ is the regular Bessel function.
Moreover, $\lap_N$ is the Laplace operator acting on functions in $\M^N$,  
then it has a discrete spectrum with eigenvalues
$\la_\al^2$ and eigenvectors $\phi_\al$.
We have set $\nu_\al^2=\la_\al^2+\rho_N^2$, 
where $\rho_N=(N-1)/2$. 
With regard to the behaviour near the conical singularities, 
the compact case is similar to the non compact one,  
then we shall approximate it with the latter
(for which we know the spectrum), but with the restriction $0<r<R$.

For the diagonal kernel of a generic operator  $F(L_D)$ one has 
\cite{chee83-18-575}
\beq
F(r,\vec \tau|L_D)=
\sum_\al\int_{0}^{\infty}F(\la^2)
|\psi_{\la\al}|^2\:\la\:d\la
\:.\eeq
Integrating on the transverse coordinates we obtain the reduced trace 
$F(r|L_D)$ on $\M^N$ in the form
\beq
F(r|L_D)=\frac1{r^D}\sum_\al
\int_{0}^{\infty}F(\fr{\la^2}{r^2})\:
J^2_{\nu_\al}(\la)\:\la\:d\la
\:.\label{spec}\eeq
As it stands, such an expression is only formal, since 
the series and the integral could not be convergent.

Since we are mainly interested in the $\zeta$-function,  we choose
$F(L_D)=L_D^{-s}$ and, using Eq.~(\ref{spec}), we formally have
\beq
\ze(s;r|L_D)&=&r^{2s-D}\sum_\al
\int_{0}^{\infty}\la^{1-2s}\:J^2_{\nu_\al}(\la)\:d\la
\:.\label{vb}\eeq
The integration over $\la$ can be performed providing that 
$\frac12<\Re s<\Re \nu_\al+1$, while the series in $\al$ 
converges if $\Re s>D/2$. These restrictions have a non vanishing 
intersection for any $\al$ if $\nu_\al>\rho_N$.
If such a condition is not satisfied (this is the more common case), 
one has to treat separately low and high eigenvalues, considering
$\nu_\al\leq\nu_{\tilde\al}\leq\rho_N$ and 
$\nu_\al>\nu_{\tilde\al}$ as in the paper of Cheeger \cite{chee83-18-575}. 
Thus we write
\beq
\ze_<(s;r|L_D)&=&r^{2s-D}\:\frac{\Ga(s-\fr12)}{\sqrt{4\pi}\Ga(s)}
\sum_{\al\leq\tilde\al}\frac{\Ga(\nu_\al-s+1)}{\Ga(\nu_\al+s)}
\:,\nn\\
\ze_>(s;r|L_D)&=&r^{2s-D}\:\frac{\Ga(s-\fr12)}{\sqrt{4\pi}\Ga(s)}
\sum_{\al>\tilde\al}\frac{\Ga(\nu_\al-s+1)}{\Ga(\nu_\al+s)}
\:,\eeq
and, after the analytic continuation, we may
define the $\zeta$-function as the sum of $\ze_<$ and $\ze_>$.

We anticipate the final result, which reads
\beq
\ze(s;r|L_D)&=&\ze_<(s;r|L_D)+\ze_>(s;r|L_D)=
r^{2s-D}\:\frac{\Ga(s-\fr12)}{\sqrt{4\pi}\Ga(s)}
\aq G(s)+\ca N\Ga(1-s)\cq
\:,\nn\\
&=&r^{2s-D}\:\frac{\Ga(s-\fr12)}{\sqrt{4\pi}\Ga(s)}\aq
\sum_{j=0}^{\ii}c_j(s)\ze(s+j-\fr12|L_N)+F(s)
+ \ca N \Ga(1-s)\cq
\:,\label{zsLD}\eeq
where $L_N=-\lap_N+\rho_N^2$, $\ca N$ represents the number of zero modes
of $L_N$, while the properties of $G(s)$, 
$F(s)$ and $c_j(s)$ will be studied in 
some detail in the following. As we shall see,
$F(0)=0$, $c_0(s)=1$, while $c_j(0)=0$ for any $j>0$. 
Furthermore, the meromorphic structure of $\ze(s|L_N)$ is given 
by Seeley theorem \cite{seel67-10-172} and reads 
\beq
\ze(s|L_N)=\frac{1}{\Ga(s)}\aq 
\sum_{n=0}^{\ii}\frac{K_n(L_N)}{s-\frac{N-n}2}
+\mbox{ analytic term}\cq\:,
\eeq 
$K_n(L_N)$ being the well known Seeley-De Witt coefficients.
As a consequence, from Eq.~(\ref{zsLD}) we immediately get
\beq
\ze(0;r|L_D)=r^{-D}\:\frac{K_D(L_N)}{\sqrt{4\pi}}
\:.\eeq
In the non compact case the integration of the latter equation over $r$, 
with the measure $r^N\:dr$ requires two cutoffs for small and large 
$r$ respectively. For the  trace at $ s=0$ one gets
\beq
\ze(0|L_D)=\int_{\ep}^{\La}\ze_a(0;r|L_D)r^N\:dr=
\frac{K_{D}(L_N)}{\sqrt{4\pi}}\:\ln\frac{\La}{\ep}
\:,\eeq
namely it is logarithmically divergent.
The same result can be obtained first integrating $\ze(s;r|L_D)$ with 
respect to $r$ and then taking the limit $s\to0$.
However, in the compact case we may perform the integration considering 
$\Re s$ sufficiently large, in order to have the convergence at $r=0$. 
Thus, near $s=0$ we have 
\beq
\ze(s|L_D)&=&\frac{R^{2s}\Ga(s-\fr12)}{2\sqrt{4\pi}\Ga(s+1)}
\aq G(s)+\ca N\Ga(1-s)\cq
\nn\\
&=&\frac{K_D(L_N)}{2\sqrt{4\pi}}\:\frac1s
+\frac{K_{D}(L_N)}{\sqrt{4\pi}}\:\ln R-\frac{\ca N}2+O(s)\:.
\label{zeDs}\eeq
This result must be compared with Ref.~\cite{bord96u-82}, where the special case 
in which $M^N$ is a sphere of radius $a$ has been analyzed. 
This is our main result. In the compact generalised cone the $\zeta$-function 
of the Laplace operator may have a pole at $s=0$, in contrast with the 
Minakshisundaram theorem \cite{mina49-1-242}, 
which states that the $\zeta$-function is 
regular in the smooth compact case.  In the non compact case one gets 
a logarithmic divergence. 
Furthermore, if the base of the generalised cone $\M^N$ 
is an even-dimensional, smooth manifold without boundary, 
then $K_{D}(L_N)=0$ and the $\zeta$-function has the usual 
meromorphic structure. 
This is also trivially true for the pure (2-dimensional) cone case, 
since  the base is a flat manifold.
If $\M^N$ has boundary, then the pole at $s=0$ is always present. 

Alternatively, the singularity at $s=0$ of the $\zeta$-function could be 
traced back from the asymptotics of the heat-kernel trace. 
In fact one has
\beq
\Tr e^{-tL_D}=\frac{1}{2\pi i}
\int_{\Re s>D/2}\Ga(s)t^{-s}\ze(s|L_D)\:ds
\:.\label{hk}\eeq 
Shifting the vertical contour to the left one gets the asymptotic 
expansion for short $t$ and in particular the pole of the second 
order at $s=0$ gives rise to a logarithmic term in $t$ 
proportional to $K_D(L_N)$, 
in agreement with the Cheeger's result \cite{chee83-18-575}. 
Conversely, if one has a   
logarithmic term in $t$ in the heat-kernel expansion, the 
presence of a simple pole at $s=0$ in the $\zeta$-function directly follows 
(see for example Ref.~\cite{bord96u-82}).
In a different setting, this happens if one is dealing with a scalar 
massive field on the hyperbolic space-time $R\times H^3/\Ga$, 
$H^3/\Ga$ being a non compact hyperbolic manifold with finite volume 
\cite{byts96u-377}.

Now we outline the procedure leading to the analytic continuation of the  
$\zeta$-function at $s=0$. Eq.~(\ref{zsLD}) implies that it is sufficient 
to deal with function $G(s)$. Thus the starting point is the series
\beq
G(s)=\sum_{\nu_\al\neq0}\frac{\Ga(\nu_\al-s+1)}{\Ga(\nu_\al+s)}
\:,\label{Gas}\eeq 
which has been introduced in Ref.~\cite{chee83-18-575}.
In Eq.~(\ref{Gas}) $\nu^2_\al=\la_\al^2+\rho^2_N$. We recall that 
$L_N=-\lap_N+\rho^2_N$, $\lap_N$ being the Laplace operator acting on 
functions in $\M^N$ and $\rho_N=(N-1)/2$. 
With regard to the convergence of the latter series we observe that
for $\nu\to\ii$ one has the asymptotic expansion
\beq
\frac{\Ga(\nu-s+1)}{\Ga(\nu+s)}\sim
\nu^{1-2s}\sum_{j=0}^{\ii}c_j(s)\nu^{-2j}
\label{asymp}\eeq
and from Weyl's theorem the degeneracy of the eingenvalues behaves asymptotically as $\nu_\al^{N}$. 
As a result, the series in 
Eq.~(\ref{Gas}) is convergent for $\Re s>(N+1)/2$. 
The $c_j(s)$ coefficients in Eq.~(\ref{asymp}) are  computable 
using the asymptotic expansion of $\Ga(z)$ for large $z$, namely
\beq
\Ga(z)\sim\sqrt{\frac{2\pi}z}e^{-z+z\ln z+B(z)}
\:,\hs B(z)=\sum_{j=0}^{\ii}\frac{B_{2j}z^{1-2j}}{2j(2j-1)}
\:,\nn\eeq
$B_j$ being the Bernoulli numbers.
It it easy to see that the function 
$\frac{\Ga(\nu-s+1)}{\Ga(\nu+s)}$ for any $s=-n/2$ ($n=-1,0,1,2,...$) 
is effectively a polynomial of order $\nu^{n+1}$, in fact
\beq 
\frac{\Ga(\nu+n/2+1)}{\Ga(\nu-n/2)}=\ag\matrix
{\nu\aq\nu^2-1\cq
\aq\nu^2-2^2\cq\cdots
\aq\nu^2-\at\frac n2\ct^2\cq\:,&n=0,2,4,...\cr
\aq\nu^2-\at\frac12\ct^2\cq
\aq\nu^2-\at\frac32\ct^2\cq\cdots
\aq\nu^2-\at\frac n2\ct^2\cq\:,&n=-1,1,3,...\cr}
\cp\label{GammaP}\eeq
Then it follows that $c_j(-n/2)$ must vanish for all $j>(n+1)/2$
(they have a simple zero at $s=-n/2$, $n=-1,0,1,2,...$). 
One can directly verify that also $c_j(1)=0$ for all $j>0$.
The first coefficients can be  computed and read
\beq
c_0(s)=1\:,\hs
c_1(s)=\frac{s(s-1/2)(s-1)}3;\hs 
c_2(s)=\frac{s(s^2-1/4)(s^2-1)(s-6/5)}{18}
\:.\label{GjCoeff}\eeq

It has to be noted that $G(s)$ has a simple poles at $s=\frac{N+1}2$
and is certainly analytic for $\Re s>\frac{N+1}2$.
In order to make the analytic continuation for any $s$ we define
\beq
f_{\tilde n}(\nu,s)=\frac{\Ga(\nu-s+1)}{\Ga(\nu+s)}
-\sum_{j=0}^{\aq\frac{\tilde n}2\cq+1}c_j(s)\nu^{1-2s-2j}
\:\:\sim\:\:c_{\aq\frac{\tilde n}2\cq+2}(s)
\nu^{-\at2s+2\aq\frac{\tilde n}2\cq+3\ct}
\:,\nn\eeq
where $\aq\frac{\tilde n}2\cq$ represents the integer part 
of $\frac{\tilde n}2$. 
For $\Re s>\frac{N+1}2$ we have
\beq
G(s)=\sum_{j=0}^{\aq\frac{\tilde n}2\cq+1}
c_j(s)\:\ze(s+j-\fr12|L_N)
+F(s)\:,\hs F(s)=\sum_{\al}f_{\tilde n}(\nu_\al,s)\:.
\label{Gs}\eeq
Now, the right hand side of the latter equation has meaning for
$\Re s>\frac{N-3}2-\aq\frac{\tilde n}2\cq$ and so we have 
obtained the analytic continuation we were looking for.
It is interesting to observe that the functions $f_{\tilde n}(\nu,s)$, 
for a sufficiently large $\tilde n$, are identically vanishing 
at $s=1,1/2,0,-1/2,-1,...$ and as a consequence also $F(s)$ is vanishing 
in all that points.
This fact permits us to compute 
the behaviour of  $G(s)$ in a neighbourhood of the half-integer points
$s=-n/2$ ($n=-2,-1,0,1,2,...$). In particular, near $s=0$ one obtains
\beq
G(s)&=&\frac{K_{N+1}(L_N)}{\Ga(-1/2)s}+\tilde\ze(-\fr12|L_N)
\nn\\&&\hs\hs
+\left|\frac{K_{N-1}(L_N)}{\Ga(1/2)}\frac{c_1(s)}s+
\frac{K_{N-3}(L_N)}{\Ga(3/2)}\frac{c_2(s)}s+...\right|_{s=0}
\:,\eeq
$K_n(L_N)$ being the spectral coefficients and $\tilde\ze(s_0|L_N)$ the 
finite part of the $\zeta$-function at $s_0$.

If $\M^N$ is a smooth manifold without boundary, 
then all spectral coefficients with odd $n$ are vanishing. Thus, 
for even $N$ (odd $D$), it follows that $G(0)=\ze(-1/2|L_N)$, while
for odd $N$ (even $D$), the first term in the latter equation 
gives rise to the ``anomalous'' divergent contribution in Eq.~(\ref{zeDs}). 

As a simple example let us  consider the
pure cone with deficit angle $2\pi-\be$. 
The eigenvalues of $L_1$ are $\la_\al=\nu_\al^2=(2\pi\al/\be)^2$, $\al\in\Z$
and moreover $\rho_1$ is vanishing as well as 
all $K_n$, but $K_0=\frac{\be}{\sqrt{4\pi}}$.
Disregarding the null eigenvalue we get 
\beq 
\ze(s|L_1)=\at \frac{2\pi}{\be}\ct^{-2s}
\sum_{\al\in\Z,\al\neq0}\al^{-2s}
=2\at\frac{\be}{2\pi}\ct^{2s}\ze_R(2s)\:,
\eeq
from which the well known result
\beq
G(0)=\frac16\at\frac{\be}{2\pi}-\frac{2\pi}{\be}\ct\:.
\eeq
directly follows. Here $\ze_R$ is the usual Riemann's zeta-function.
Note that in Ref.~\cite{zerb96-54-2699} the function $G(s)$ is defined in a 
slightly different way.

We conclude with some remarks. 
In this letter the analytic properties of the $\zeta$-function 
related to the
Laplace operator on a generalised cone $\R^+\X\M^N$ have been 
investigated using the Cheeger's approach and explicit expressions for 
it have been obtained.
We have shown that in the compact case, 
the $\zeta$-function of the Laplace operator  for a
minimally coupled massless scalar field turns out to be 
singular at the origin. 
As a consequence, since the one-loop effective action 
in the  $\zeta$-function regularisation approach
is formally given by $-\zeta'(0|L_D)/2$, 
looking at Eq.~(\ref{zeDs}) one sees that a further subtraction 
is required in order to remove the singularity at $s=0$. 
This singularity is proportional to the spectral coefficient 
$K_D(-\lap_N+\rho^2_N)$,
thus in principle, the nature of the counterterm is known. 
It has to be noted that if the base of the generalised cone 
is a manifold with boundary, also in the odd-$D$ dimensional 
case the singularity at $s=0$ of the $\zeta$-function is present.

\end{document}